\documentclass[review]{elsarticle}
\usepackage{color}
\usepackage{amssymb}
\usepackage{lineno,hyperref}
\journal{Journal of \LaTeX\ Templates}

\newcommand{\jpsi}{\mathrm{J/}\psi}
\newcommand{\Rz}{\rho^{0}}
\newcommand{\Uos}{\Upsilon(1S)}

\newcommand{\Wgp}{W_{\gamma\mathrm{p}}}
\newcommand{\WgA}{W_{\gamma\mathrm{A}}}
\newcommand{\sNN}{\sqrt{{\textit s}_{\rm NN}}}

\begin{document}

\begin{frontmatter}

\title{Mass dependence of  vector meson photoproduction off protons and nuclei within the 
energy-dependent hot-spot model}

\author[CVUT]{J. Cepila}
\author[CVUT]{J. G. Contreras}
\author[CH,CVUT]{M. Krelina}
\author[KU]{J. D. Tapia Takaki}
\address[CVUT]{Faculty of Nuclear Sciences and Physical Engineering,
Czech Technical University in Prague, Czech Republic}
\address[CH]{Departamento de F\'{\i}sica, Universidad T\'ecnica Federico Santa Mar\'{\i}a;
Centro Cient\'{\i}fico-Tecnol\'ogico de Valpara\'{\i}so-CCTVal,
 Casilla 110-V, Valpara\'{\i}so, Chile}
\address[KU]{Department of Physics and Astronomy, The University of Kansas, Lawrence, KS, USA}

\begin{abstract}
We study the photoproduction of vector mesons off proton and off nuclear targets. We work  within the colour dipole model in an approach that includes subnucleon degrees of freedom, so-called hot spots, whose positions in the impact-parameter plane change event-by-event.  The key feature of our model is that the number of hot spots depends on the energy of the photon--target interaction. Predictions are presented for exclusive and dissociative production of $\Rz$, $\jpsi$, and $\Uos$ off protons, as well as for coherent and incoherent photoproduction of $\Rz$ off nuclear targets, where Xe, Au, and Pb nuclei are considered.  We find that the
mass dependence of dissociative production off protons as a function of the energy of the interaction provides a further handle to search for saturation effects at HERA, the LHC and future colliders. We also find that the coherent photonuclear production of $\Rz$ is sensitive to fluctuations in the subnucleon degrees of freedom at RHIC and LHC energies.
\end{abstract}


\end{frontmatter}

 %\linenumbers

%%%%%%%%%%%%%%%%%%%%%%%%%%%%%%%%
%% New section: 
\section{Introduction
\label{sec:Intro}}
%%%%%%%%%%%%%%%%%%%%%%%%%%%%%%%%

The diffractive photoproduction of vector mesons in high-energy interactions is
sensitive to the energy evolution of the gluon distribution of hadrons at small~$x$, as well as the distribution of gluonic degrees of freedom in the impact-parameter plane. As such, it has been extensively studied at HERA~\cite{Ivanov:2004ax,Newman:2013ada} and at the LHC~\cite{Contreras:2015dqa,Andronic:2015wma}; it is also one of the key observables to search for gluon saturation ~\cite{Gelis:2010nm,Albacete:2014fwa} in future facilities~\cite{Accardi:2012qut, AbelleiraFernandez:2012cc}.

Recently, this type of process has been used to investigate the behaviour of subnucleon degrees of freedom in the structure of hadrons. In particular,  it has been shown~\cite{Mantysaari:2016ykx,Mantysaari:2016jaz} that in a Good-Walker formalism~\cite{Good:1960ba,Miettinen:1978jb} the dissociative  photoproduction of $\jpsi$ is sensitive to the event-by-event fluctuations, in the impact-parameter plane, of so-called hot spots --- regions of high gluonic density. These studies  fixed the number of hot spots within the proton to three. The same model was then used to investigate these processes for the case of nuclear targets~\cite{Mantysaari:2017dwh}. 

In a further development, the diffractive photoproduction of $\jpsi$ was studied in a model in which the number of hot spots increases with decreasing $x$~\cite{Cepila:2016uku}. This model was also successfully extended to describe $\jpsi$ photoproduction off nuclear targets at the LHC~\cite{Cepila:2017nef}.
The energy dependence on the number of hot spots provides new signatures for saturation: in the case of production off protons, the dissociative cross section reaches a maximum after which it decreases steeply at high energies~\cite{Cepila:2016uku}; while for the production off lead the ratio of the incoherent to the coherent cross section shows a slope as a function of $x$~\cite{Cepila:2017nef}. Both behaviours occur in energy ranges accessible to the LHC. 

In this work, we extend our studies to include diffractive photoproduction of $\Rz$ and $\Uos$ vector mesons for two reasons. There are HERA, RHIC and LHC data available for these vector mesons, and we would like to explore the limits of our model. Our main result is shown in Fig.~\ref{fig:gp_VMY}: the position of the maximum of the dissociative cross section shows a dependence on the mass of the vector meson, providing a further handle to search for saturation effects.

The rest of this document is organised as follows. In the next section, we review the formalism used in this work. Section~\ref{sec:Results} reports our results and compares the predictions of our model to experimental data. In Sec.~\ref{sec:Discussion}, we discuss the results and explore its implications. Finally, in Sec.~\ref{sec:Summary} we summarise our work and findings.

%%%%%%%%%%%%%%%%%%%%%%%%%%%%%%%%
%% New section: 
\section{Review of the formalism
\label{sec:formalism}}
%%%%%%%%%%%%%%%%%%%%%%%%%%%%%%%%

Here, we give an overview of our model. More details can be found in~\cite{Cepila:2016uku, Cepila:2017nef}. With two exceptions, discussed in Sec.~\ref{sec:Discussion}, the values of all parameters are the same as in these two references.

The diffractive photoproduction of vector mesons proceeds,  within  the  colour dipole model~\cite{Mueller:1989st,Nikolaev:1990ja},  in three steps: the fluctuation of the incoming photon into a quark-antiquark pair, the interaction of this pair with the target --- here we considered a  dipole--proton (dp) or a dipole--nucleus (dA) interaction --- and the formation of a vector meson.  In this approach, the amplitude for photon--proton interactions can be written as follows (e.g.,~\cite{Kowalski:2006hc}),

\begin{equation}
A(x,Q^2,\vec{\Delta})_{T,L} = i\int d\vec{r}\int^1_0\frac{dz}{4\pi}
(\Psi^*\Psi_{\rm VM})_{T,L} \int d\vec{b}\;
e^{-i(\vec{b}-(1-z)\vec{r})\cdot\vec{\Delta}}\frac{d\sigma_{\rm dp}}{d\vec{b}},
\label{eq:Amplitude}
\end{equation}
where $-t = \vec{\Delta}^2$ is the momentum transferred at the target vertex, and $\Psi$ is the wave function of a virtual photon, of virtuality $Q^2$, fluctuating into a quark-antiquark colour dipole.
For all vector mesons the wave function  $\Psi_{\rm VM}$ is obtained from the 
boosted-Gaussian model~\cite{Nemchik:1994fp,Nemchik:1996cw}, with the numerical values of the parameters  as in~\cite{Kowalski:2006hc}.  The dipole is described by   the transverse distance between the quark and the antiquark, $\vec{r}$, and by the fraction of the longitudinal momentum of the dipole carried by the quark, $z$. The
impact parameter is denoted by $\vec{b}$. Finally,  $T$ and $L$ refer to the contribution of the transversal and longitudinal degrees of freedom of the virtual photon, respectively. 

In our model, the $\vec{r}$ and $\vec{b}$ dependences of the dipole--proton interaction are factorised as
\begin{equation}
\frac{d\sigma_{\rm dp}}{d\vec{b}} = 2 \sigma_0N(x,r)T(\vec{b}).
\label{eq:b-depSigDip}
\end{equation}
Here, $r$ is the magnitude of $\vec{r}$. The value of $\sigma_0$ is fixed by the value of the proton profile in impact parameter space through the value of $B_p$ introduced below. We chose the form of $N(x,r)$ given by the model of Golec-Biernat and Wusthoff~\cite{GolecBiernat:1998js}:
\begin{equation}
 N(x,r) = \left( 1-e^{-r^2Q^2_s(x)/4}\right), \ \ \ Q^2_s(x) = Q^2_0(x_0/x)^\lambda,
\end{equation}
with the saturation scale, $Q_s(x)$,  given by the parameters $\lambda$, $x_0$ and $Q^2_0$.
For the proton profile in impact-parameter space we use the sum of $N_{hs}$ hot spots defined by Gaussian distributions of width $B_{hs}$, and positions obtained from another Gaussian distribution centred at the origin and having a width $B_p$:
\begin{equation}
T(\vec{b}) = \frac{1}{N_{hs}}\sum^{N_{hs}}_{i=1}T_{hs}(\vec{b}-\vec{b_i}),
\end{equation}
with
\begin{equation}
T_{hs}(\vec{b}-\vec{b_i}) = \frac{1}{2\pi B_{hs}}e^{-\frac{(\vec{b}-\vec{b_i})^2}{2B_{hs}}}.
\end{equation}

For nuclei we use the same form for the amplitude as in Eq.~\ref{eq:Amplitude}, but using  a
 dipole--nucleus cross section, which was obtained from the dipole--proton cross section defined above in two approaches~\cite{Cepila:2017nef}.  One based on a Glauber-Gribov methodology proposed in~\cite{Armesto:2002ny}, denoted GG, and one based on geometric-scaling ideas as proposed 
 in~\cite{Armesto:2004ud}, denoted GS.

For the GG case, we have \begin{equation}
\frac{d\sigma_{\rm dA}}{d\vec{b}} = 2\left[
1-\exp\left(-\frac{1}{2}
\sigma_0N(x,r)T_{\rm A}(\vec{b})\right)\right],
\label{eq:GG}
\end{equation}
while for GS
\begin{equation}
\frac{d\sigma_{\rm dA}}{d\vec{b}} = 
\sigma^{\rm A}_0\left[1-\exp\left(-r^2Q^2_{A,s}(x)/4\right)\right]
T_{\rm A}(\vec{b}),
\end{equation}
with  $\sigma^{\rm A}_0$  related to the area of the target by $\sigma^{\rm A}_0=\pi R^2_{\rm A}$, where $R_{\rm A}$ is the radius of the nucleus, which we take from  the  Woods-Saxon distribution of the given nucleus. The saturation scale  of the nucleus is given by~\cite{Armesto:2004ud}
\begin{equation}
Q^2_{s,{\rm A}}(x) = Q^2_s(x) \left(\frac{{\rm A}\pi R^2_{\rm p}}{\pi R^2_{\rm A}}\right)^{\frac{1}{\delta}}.
\label{eq:QsA}
\end{equation}

The nuclear profile, $T_{\rm A}(\vec{b})$ is given either by a set of Gaussian nucleons of width $B_p$, whose centres are obtained from a Woods-Saxon distribution, or by adding to the nucleons a substructure made of hot spots as described above. Below, these cases are denoted by GG-n and GG-hs, respectively.

The evolution of the number of hot spots with energy is modelled, both for proton and nuclear targets, with
a random number, $N_{hs}$, drawn from a zero-truncated Poisson distribution, where the Poisson distribution has a mean value
\begin{equation}
\langle N_{\rm hs}(x) \rangle = p_0x^{p_1}(1+p_2\sqrt{x}),
\label{eq:Nhsx}
\end{equation}
where $p_0$, $p_1$ and $p_2$ are parameters.

Using the amplitude presented in Eq.~\ref{eq:Amplitude}, the corresponding cross sections for the photoproduction of a vector meson VM off a target T are
\begin{equation}
\left.
\frac{d\sigma(\gamma {\rm T}\rightarrow {\rm VM\; T})}{dt}
\right|_{T,L} =
 \frac{(R^{T,L}_g)^2}{16\pi}
 \left| 
\left< A(x,Q^2,\vec{\Delta})_{T,L}\right>
\right|^2,
\label{eq:ExclXS}
\end{equation}
for the exclusive or coherent processes, and
\begin{equation}
\left.
\frac{d\sigma(\gamma {\rm T}\rightarrow {\rm VM\;} Y)}{dt} 
\right|_{T,L}= \frac{(R^{T,L}_g)^2}{16\pi}
\left(\left<\left| 
A(x,Q^2,\vec{\Delta})_{T,L}
\right|^2\right>
-
\left| 
\left< A(x,Q^2,\vec{\Delta})_{T,L}
\right>\right|^2\right),
\label{eq:DissXS}
\end{equation}
 for dissociative or incoherent production, where $Y$ represents the dissociative state. $R^{T,L}_g$ is the skewedness correction~\cite{Shuvaev:1999ce}.
The total cross section at a given $t$ is the sum of the  $T$ and $L$ contributions. The integral over $t$ yields the cross section at a given energy.

%%%%%%%%%%%%%%%%%%%%%%%%%%%%%%%%
%% New section: 
\section{Results: comparison to data and predictions
\label{sec:Results}}
%%%%%%%%%%%%%%%%%%%%%%%%%%%%%%%%

Using the formalism reviewed in the previous section we compute the cross sections given in Eqs.~(\ref{eq:ExclXS}) and (\ref{eq:DissXS}) for the photoproduction of a vector meson off a proton using 10000 configurations of the profile function at each value of $x$. As it is customary, we relate the value of $x$ to the centre-of-mass energy of the photon--proton system, $\Wgp$, by $x\equiv M^2_{\rm VM}/\Wgp^2$.  

\begin{figure}[!t]
\centering % \begin{center}/\end{center} takes some additional vertical space
\includegraphics[width=0.98\textwidth]{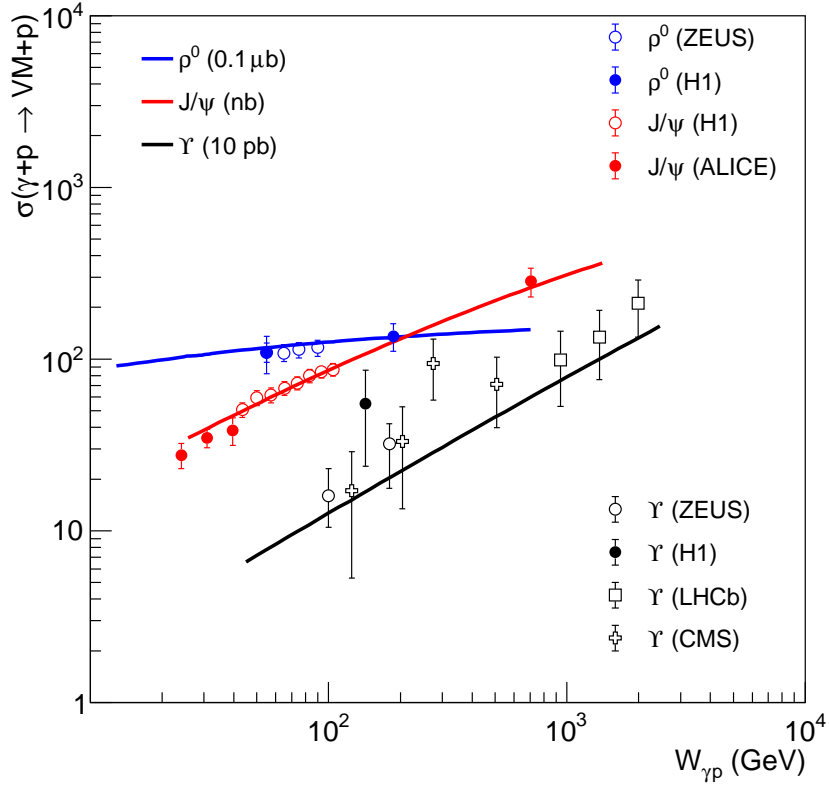}
\caption{
\label{fig:gp_VMp} 
Energy dependence for $\Rz$, $\jpsi$ and $\Uos$ exclusive photoproduction off protons as predicted by our model (solid lines) compared to measurements from H1~\cite{Aid:1996bs,Alexa:2013xxa,Adloff:2000vm}, ZEUS~\cite{Breitweg:1997ed,Chekanov:2009zz}, ALICE~\cite{TheALICE:2014dwa}, LHCb~\cite{Aaij:2015kea} and CMS~\cite{CMS-PAS-FSQ-13-009}. Note that the different cross sections  are displayed  in different units.
}
\end{figure}

Figure \ref{fig:gp_VMp} shows the cross section for the exclusive production of a vector meson off a proton as a function of $\Wgp$ for the three particles studied here: $\Rz$, $\jpsi$ and $\Uos$. The predictions of the model are compared to the available data from HERA and the LHC. For the case of the $\Rz$ we compare to data from H1~\cite{Aid:1996bs} and ZEUS~\cite{Breitweg:1997ed}, for the $\jpsi$ to data from H1~\cite{Alexa:2013xxa} and ALICE~\cite{TheALICE:2014dwa}, while for the $\Uos$ we compare to data from H1~\cite{Adloff:2000vm}, ZEUS~\cite{Chekanov:2009zz}, LHCb~\cite{Aaij:2015kea} and preliminary data from CMS~\cite{CMS-PAS-FSQ-13-009}. The cross sections vary from some 20 pb to some 100 $\mu$b, and the energy $\Wgp$ ranges from 20 GeV up to 2 TeV. In all this domain the description of data by the model is quite satisfactory.

\begin{figure}[!t]
\centering % \begin{center}/\end{center} takes some additional vertical space
\includegraphics[width=0.98\textwidth]{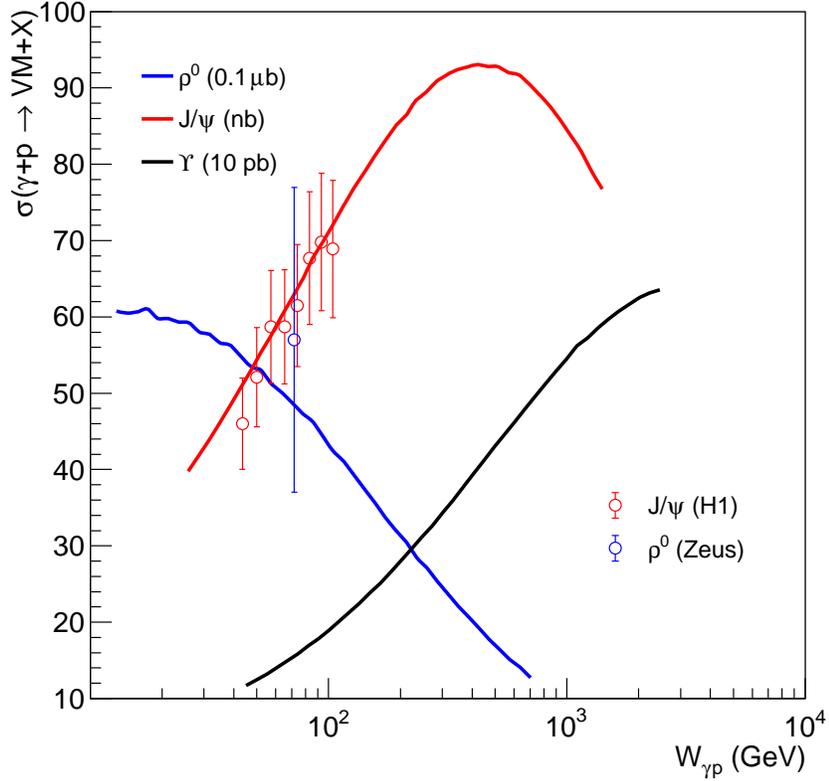}
\caption{
\label{fig:gp_VMY} 
Energy dependence for $\Rz$, $\jpsi$ and $\Uos$ dissociative photoproduction off protons as predicted by our model (solid lines) compared to measurements from H1~\cite{Alexa:2013xxa} and ZEUS~\cite{Breitweg:1997ed}. Note that the different cross sections  are displayed  in different units.
}
\end{figure}

Figure \ref{fig:gp_VMY} shows the main result of this work: the energy dependence of the cross section for the dissociative production of a vector meson off a proton as a function of $\Wgp$ for  $\Rz$, $\jpsi$ and $\Uos$. The predictions of the model are compared to data from ZEUS~\cite{Breitweg:1997ed}  and H1~\cite{Alexa:2013xxa} for the $\Rz$ and the $\jpsi$, respectively. (Note that in~\cite{Breitweg:1997ed} the exclusive cross section and the ratio of the exclusive to the dissociative cross section are quoted; the data point shown in the figure is derived from these values.) 
As already observed in~\cite{Cepila:2016uku} for the $\jpsi$, the dissociative cross section presents a maximum after which there is a steep descent.  The cross sections for $\Rz$ and $\Uos$ show the same behaviour. 
The position of the maximum is correlated with the mass of the vector meson. For $\Rz$ it is around 10-20 GeV, while for $\Uos$ it is above two TeV.

\begin{figure}[!t]
\centering % \begin{center}/\end{center} takes some additional vertical space
\includegraphics[width=0.98\textwidth]{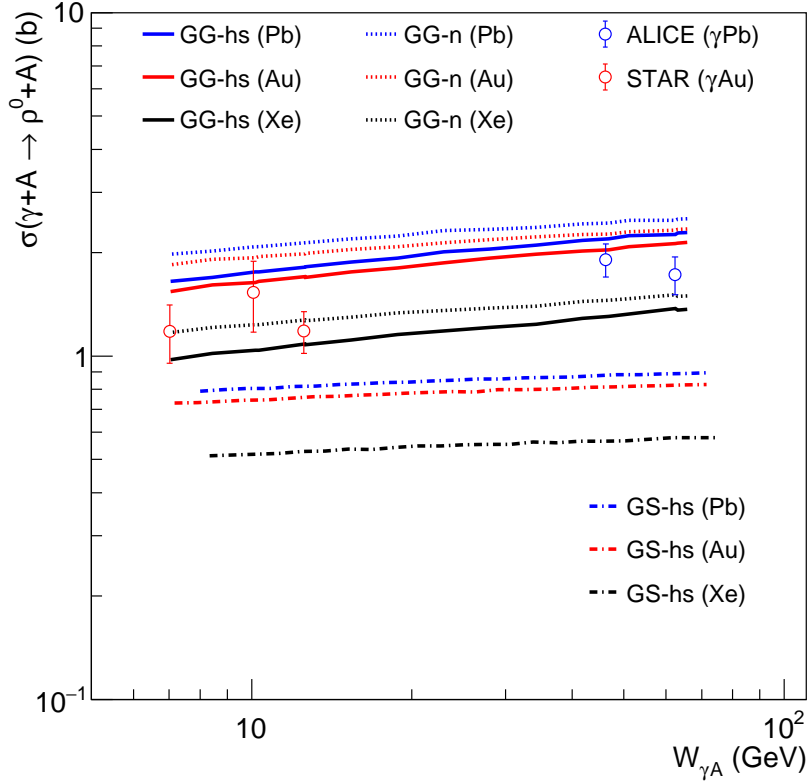}
\caption{
\label{fig:gA_VMA} 
Energy dependence for $\Rz$  coherent photoproduction off a nucleus (A) as predicted by our model. The approach based on geometric-scaling ideas, GS, is shown with dashed-dotted lines, while the Glauber-Gribov, GG, approach is shown with solid (dotted)  lines for the calculation including  hot spots (only nucleons). The model is  compared to measurements from RHIC~\cite{Adler:2002sc,Abelev:2007nb,Agakishiev:2011me} and the LHC~\cite{Adam:2015gsa,Pozdniakov:2017fwn}. 
}
\end{figure}

For the nuclear case, we study the coherent and incoherent photoproduction of $\Rz$ off the following nuclei: Xe, Au, and Pb. The nuclei are modelled with Woods-Saxon distributions. The deformation of the Xenon nucleus is  taken into account parameterising the radius of the nucleus as a function of the polar angle. For the GS (GG) approach we  only use  1000 (200) configurations, due to the increase in computational resources needed for the calculations. 

%Results for $\jpsi$ photonuclear production were reported in~\cite{Cepila:2017nef}. 
The results of our model for $\Rz$ coherent photoproduction off a nulceus (A) at a photon--nucleus centre-of-mass energy $\WgA$ are shown in Fig. \ref{fig:gA_VMA} for the GG-hs, GG-n and GS-hs cases. (Note that the results for GS-n  are the same as those from GS-hs, so they are not shown.) The predictions of the model are compared to the available data from STAR at RHIC~\cite{Adler:2002sc,Abelev:2007nb,Agakishiev:2011me} and ALICE at the LHC, both for the published data~\cite{Adam:2015gsa} as well as for the preliminary results~\cite{Pozdniakov:2017fwn}. 
The published measurements were performed at mid-rapidity with ultra-peripheral collisions of Au--Au and Pb--Pb, respectively, allowing one to extract the $\gamma$--nucleus cross section by taking into account the corresponding photon flux. 
%This was done using the Starlight Monte Carlo~\cite{Klein:2016yzr}. 
The energy dependence of the GG-hs curve is steeper than that of the GG-n. This is due to the exponential in Eq.~\ref{eq:GG} which is sensitive to the different profiles of the GG-n and GG-hs cases.  For $\WgA$ energies around 500 GeV, both profiles are so similar that both computations yield the same cross section. Both GG variantes are steeper than the prediction using the GS-hs approach.

Up to now, there are no measurements of the incoherent photonuclear production of $\Rz$, but the collaborations at RHIC and the LHC have (or will have in the future) data sets that could be used to perform such measurements. 
Table~\ref{tab:xs} presents our predictions  for the coherent and the incoherent photonuclear production of $\Rz$ at mid-rapidity in 
ultra-peripheral Pb--Pb and Xe--Xe collisions for the corresponding energies of the Run 2 at the LHC.

\begin{table}
\centering
\caption{\label{tab:xs} 
Cross sections for the coherent and the incoherent photonuclear production of $\Rz$ at mid-rapidity in ultra-peripheral Au--Au, Pb--Pb and Xe--Xe collisions at energies available at RHIC and  the LHC. The cross sections are computed in the Glauber-Gribov approach considering just nucleons (GG-n) or nucleons with hot spots (GG-hs)}
~\\
~\\
\begin{tabular}{cccc}
\hline
\hline
System& $\sNN$ (TeV)  & Coherent (mb) & Incoherent (mb) \\
& & GG-n --- GG-hs&GG-n --- GG-hs\\
\hline
Au--Au & 0.20 & 118 --- 102 & 3.6 --- 4.4\\
Pb--Pb & 2.76  & 535 --- 480& 16 --- 27\\
Pb--Pb & 5.02  & 636 --- 579& 18 --- 31\\
Xe--Xe & 5.44 & 175 --- 160& 6.5 --- 9.8\\
\hline
\hline
\end{tabular}
\end{table}

%%%%%%%%%%%%%%%%%%%%%%%%%%%%%%%%
%% New section: 
\section{Discussion
\label{sec:Discussion}}
%%%%%%%%%%%%%%%%%%%%%%%%%%%%%%%%

As mentioned above, two parameters of our model are updated for these studies. The first one is the parameter $p_2$ in Eq.~(\ref{eq:Nhsx}) which is changed from 250 to 300. The reason is that in our original model presented in~\cite{Cepila:2016uku}, we used a fixed number of hot spots at each value of $x$, while now we draw the number of hot spots from a zero-truncated Poisson distribution as described above. The extra fluctuations introduced by this new procedure induced a slight change in the results that is compensated by the new value of the $p_2$ parameter. This change does not affect at all the exclusive cross section  and as shown in Fig.~\ref{fig:gp_VMY}, neither the quality of the description of $\jpsi$ dissociative data is affected. The photonuclear cross section for $\jpsi$ is not sensitive to this change, either.

\begin{figure}[!t]
\centering % \begin{center}/\end{center} takes some additional vertical space
\includegraphics[width=0.98\textwidth]{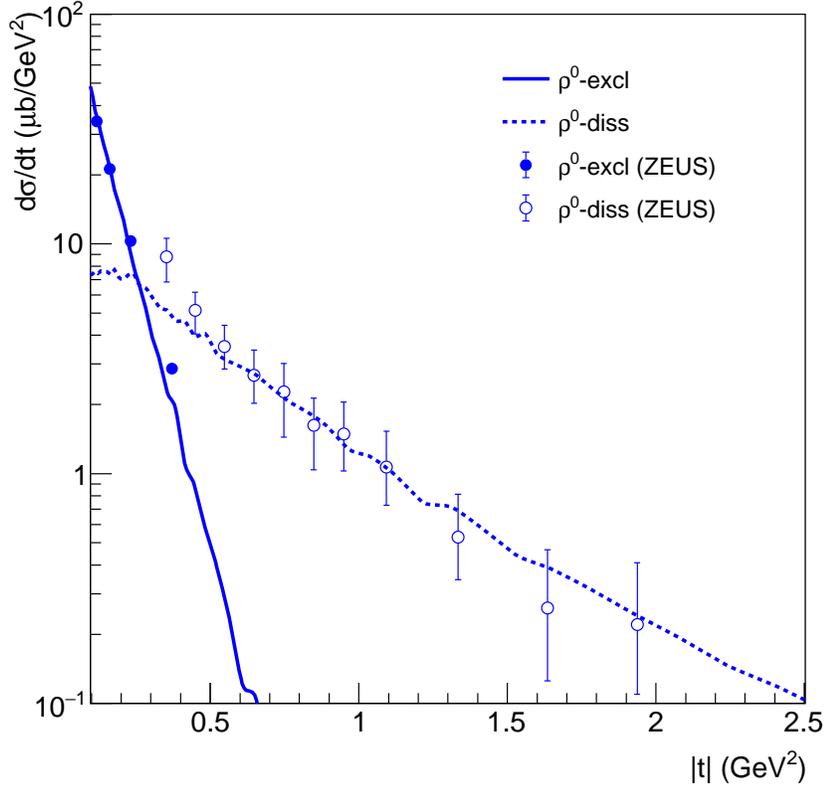}
\caption{
\label{fig:gp_rho_t} 
Dependence on $|t|$ for the exclusive (excl) and dissociative (diss) $\Rz$ photoproduction off protons at $\Wgp = 71.7$ GeV compared to data from ZEUS~\cite{Breitweg:1997ed}. 
}
\end{figure}

The value of $B_p$ for the $\jpsi$ is the same as  in our previous publications, 4.7 GeV$^{-2}$, and this same value is used for the $\Uos$ predictions, while for  $\Rz$ production we set it to  8.0  GeV$^{-2}$, in order to  obtain a good description of the $t$ dependence of the exclusive and dissociative cross sections for $\Rz$ production at $\Wgp = 71.7$ GeV. The agreement with data from ZEUS~\cite{Breitweg:1997ed} is shown in Fig.~\ref{fig:gp_rho_t}. This value of $B_p$ is in line with those values measured experimentally, as depicted in the left panel of Fig. 22 of~\cite{Aaron:2009xp}.

The applicability of the formalism expressed in Eq~\ref{eq:Amplitude} to photoproduction of $\Rz$ mesons at small $t$ has to be taken with care. The mass of the $\Rz$ may not provide a scale large enough for perturbative calculations. The only other potentially large scale which is present in the problem is the saturation scale, but it is not clear to which extend it can justify the validity of the model here. Nonetheless, we note that other authors have used the dipole colour model to perform similar computations, e.g.~\cite{Forshaw:2003ki,Rezaeian:2013tka,Armesto:2014sma}. The success of those calculations, and the agreement of our model with data shown in the previous section, points to the possibility that higher order or non-perturbative effects are, in this case, successfully absorbed in the parameters of the model.

 The GS-hs prescription yields predictions for coherent $\Rz$ photonuclear production  which are about a factor of two lower than the measurements, as shown in Fig.~\ref{fig:gA_VMA}. This seems to suggest that the scaling given by Eq.~(\ref{eq:QsA}), which gives a good description of the $\jpsi$ data~\cite{Cepila:2017nef}, is not valid for the small  scales present in $\Rz$ photoproduction. On the other hand, the GG prescription overshoots systematically the measured data by around one and a half sigmas. Nonetheless, there is an interesting observation: the inclusion of subnuclear degrees of freedom introduces an extra energy dependence to the cross sections, which can be seen in Fig.~\ref{fig:gA_VMA} by comparing the solid and dotted lines.  At the same time, the current precision of data does not allow one to conclude which model is favoured, GG-hs or GG-n. 
 Recently, the LHC produced Xe--Xe collisions, and towards the end of 2018 it will again produce  Pb--Pb collisions. It is expected that these data taking periods will yield large  samples of photonuclear produced  $\Rz$, which would allow for more precise measurements shedding new light on this issue.

The main result of this work is depicted in Fig.~\ref{fig:gp_VMY}. The dissociative cross section is related in our formalism to the variance over configurations. The fact that the dissociative cross section falls down with energy, that is, with an increasing number of hot spots, indicates that at some point there are so many hot spots that all configurations look similar and the variance decreases. This means that the hot spots saturate the transverse area of the target, making it look more and more like a black disk. The energy at which this happens decreases with decreasing mass  of the vector meson. 

Up to now, there are no data for the energy dependence of  dissociative production of $\Rz$ nor of $\Uos$. For the $\Rz$ one could imagine that a reprocessing of HERA data could be used to perform this measurement. Furthermore, such data could also be measured at the LHC in p--Pb ultra-peripheral collisions, using  forward detectors  to tag the presence of dissociation.
Note that for the energy range accessible at HERA and the LHC, our model predicts that the dissociative cross section for $\Rz$ will decrease, that for $\jpsi$ will increase, reach a maximum and decrease, and that of the $\Uos$ will only increase. This pattern is striking  and the observation of such a behaviour would be a strong indication of saturation effects.

The last point we want to comment on is about the limits of our model. The dipole--target cross sections encode the perturbative QCD knowledge that goes into the  model. We use relatively simple, but QCD inspired, forms for these cross sections.
We want to emphasise that the knowledge we want to extract from our model is  the broad behaviour of the cross sections. The agreement with data is comforting, but it should not be overemphasised. Nonetheless, the values of the parameters used in the model cannot be changed arbitrarily. They are fixed by physics arguments and by the comparison to a subset of the available data. Given the simplicity of the model, there is very little freedom to change the value of the parameters and still conserve a reasonable agreement with data. In this context, the pattern observed in  Fig.~\ref{fig:gp_VMY} is a solid prediction. 

%%%%%%%%%%%%%%%%%%%%%%%%%%%%%%%%
%% New section: 
\section{Summary and outlook
\label{sec:Summary}}
%%%%%%%%%%%%%%%%%%%%%%%%%%%%%%%%

We reported on the photoproduction of vector mesons off proton and off nuclear targets  within the colour dipole model in an approach that includes an energy-dependent hot-spot structure of the target in the impact-parameter plane. Predictions are presented for exclusive and dissociative production of $\Rz$, $\jpsi$ and $\Uos$ off protons, as well as for coherent and incoherent  photoproduction of $\Rz$ off nuclear targets, where Xe, Au, and Pb nuclei were considered.  

For coherent $\Rz$ photonuclear production we find that in a Glauber-Gribov approach the steepness of the energy dependence changes when introducing hot spots, with respect to the case where nuclei are formed just by nucleons. 

Our main finding is that there is a
mass dependence of dissociative production off protons: the cross section for this process initially increases with $\Wgp$, reaches a maximum and then decreases steeply; the position of the maxima depends on the mass of the vector meson. In our model, the decrease of the cross section is driven by saturation effects and the mass ordering of this behaviour offers  a further handle to search for the onset of gluon saturation in the proton.

The effects predicted here occur in an energy range which is accessible with HERA, RHIC and LHC data.
New data on photonuclear production is  expected from the LHC in the near future; these data may shed new light to guide our understanding of these processes. The dissociative production of vector mesons  could be tagged at the LHC in p--Pb interactions using  forward detectors as it was done at HERA. Such data could potentially be used to  test our predictions.

%%%%%%%%%%%%%%%%%%%%%%%%%%%%%%%%
%% New section: 
\section*{Acknowledgements}
%%%%%%%%%%%%%%%%%%%%%%%%%%%%%%%%

We would like to thank J. Nystrand for making available to us the experimental data shown in Fig.~\ref{fig:gA_VMA}.
This work was partially supported by grants 18-07880S of the Czech Science Foundation, 
LTC17038 of the INTER-EXCELLENCE program at the Ministry of Education, Youth and Sports of the Czech Republic, and by grants Conicyt PIA/ACT 1406 (Chile) and Conicyt PIA/Basal FB0821 (Chile).
Access to computing and storage facilities of the National Grid Infrastructure MetaCentrum provided under the programme CESNET LM2015042 of the Czech Republic is greatly appreciated.

%%%%%%%%%%%%%%%%%%%%%%%%%%%%%%%%
%% New section: 
\section*{References}
%%%%%%%%%%%%%%%%%%%%%%%%%%%%%%%%

\end{document}